\documentclass[10pt,twocolumn,twoside]{osajnl}
\usepackage{graphicx}
\usepackage{nccmath}
\usepackage{amsmath}
\usepackage{amssymb}
\usepackage{graphicx}
\usepackage{dcolumn}
\usepackage{bm}
\usepackage{xcolor}
\usepackage{caption}
\usepackage{subfigure}
\usepackage{float}
\usepackage{lineno}
\usepackage{appendix}
\usepackage{hyperref}
\usepackage{upgreek}
\usepackage{cuted}
\usepackage{amsmath,lipsum}
\usepackage{cuted}
\usepackage{multicol}

\makeatletter
\renewcommand{\ps@headings}{}
\renewcommand{\ps@plain}{}
\makeatother
\providecommand{\journalref}{}
\providecommand{\journallongtype}{}
\providecommand{\journalname}{}

\providecommand{\received}[1]{}
\providecommand{\accepted}[1]{}
\providecommand{\published}[1]{}
\providecommand{\doi}[1]{}
\providecommand{\setprjcopyright}{}
\providecommand{\address}[1]{}
\providecommand{\authorref}[1]{}
\providecommand{\email}[1]{}
\definecolor{color1}{RGB}{0,102,153}
\definecolor{color2}{RGB}{204,0,0}
\usepackage{fancyhdr}
\title{Research and simulation of analytical polarization control enabled by optical computing on an integrated photonics chip}

\author[1]{Xueying Ren}
\author[2]{Meinan Guo}
\author[1,3,*]{Xuyang Wang}
\author[4,†]{Bailin Shen}
\author[5]{Lingyan Zhang}
\author[5]{Minyue Yang}
\author[5]{Nannan Ning}
\author[5]{Jiaxin Huang}
\author[2]{Lv Lv}
\author[6]{Jun Zou}
\author[1,3,‡ ]{Yongmin Li}

\affil[1]{State Key Laboratory of Quantum Optics Technologies and Devices, Institute of Opto-Electronics, Shanxi University, Taiyuan 030006, China}
\affil[2]{College of Physics and Electronic Engineering, Shanxi University, Taiyuan 030006, China}
\affil[3]{Collaborative Innovation Center of Extreme Optics, Shanxi University, Taiyuan 030006, China}
\affil[4]{Beijing Xicheng Photonics Techlonogy Co.,Ltd., Beijing 100176, China}
\affil[5]{ZTE Photonics Technology Co., Ltd., Nanjing 210012, China}
\affil[6]{ZJU-Hangzhou Global Scientific and Technological Innovation Center, Zhejiang University, Hangzhou 311215, China}

\affil[*]{Corresponding author: wangxuyang@sxu.edu.cn, shen.bailin@zphotonics.com.cn, yongmin@sxu.edu.cn}

\begin{abstract}
Dynamic polarization controllers are key devices with broad applications in many fields. However, most on-chip polarization controllers still rely on traditional blind-search methods, whereas analytical optical-computing approaches remain insufficiently explored, particularly with respect to calibration and endless polarization control. With the accurate relative phase of Mach-Zehnder interferometer (MZI) being fully controllable on an integrated photonics chip, we present an analytical polarization control (APC) method using four phase shifters and optical computing, eliminating the need for the traditional inefficient blind-search procedure. The basic structures and operations of APC are clarified. The proposed calibration method and endless control method enable continuous APC while compensating for phase differences within the MZI structures.  We simulate the influence of the endless control unit on polarization control and quantify the effect of the fourth phase difference on the output extinction ratio.  With the fourth phase shifter, the phase difference encountered during Stokes vector measurement can be effectively compensated, and rotations around all three axes on the Poincaré sphere can be realized. These results establish a practical APC architecture based on optical computing for photonics chips. The proposed APC methods, combined with a FPGA-based hardware acceleration, will enable high speed on-chip polarization controllers.
\end{abstract} 

\setboolean{displaycopyright}{false}

\begin{document}
\makeatletter
\renewcommand{\setprjcopyright}{}
\renewcommand{\published}[1]{}
\renewcommand{\doi}[1]{}
\renewcommand{\@doi}{}
\def\@publishednote{}
\makeatother
\maketitle

\section{Introduction}
Dynamic polarization controllers (DPCs) can transform any input polarization state into a desired polarization state thereby mitigating polarization impairments caused by both internal and external induced birefringence. They have broad applications in optical communication \cite{1,2,3,yusiyuan}, optical imaging \cite{4}, biomedicine \cite{5}, data-center connectivity \cite{6,7}, optical sensing and quantum technologies \cite{8,9,10,11,12,13,14,15,16,17,18,19,20,21,22,23,Wang2022,Shao2025,Chen2025,Liu2022,Zheng2026,Zhang2026,Li2024,FanYuan2021,Su2023,Wang2021,Jin2025}.With the rapid development of optical communication and large-scale optical integration technologies, it is imperative to implement the traditional functions of optical polarization processing on an integrated platform. In recent years, several promising integrated optical polarization controller platforms have been reported \cite{6,7,27,28,29,30,31,32,33}. Mainstream integrated DPCs are implemented on two primary platforms: thin-film lithium niobate (TFLN) and silicon. By exploiting the ultrafast electro-optic response of the lithium niobate platform, TFLN-DPCs have achieved tracking speeds of up to 100 krad/s \cite{31}. In 2024, a high-speed automatic polarization controller with speed of 20 krad/s was demonstrated on silicon, incorporating innovative thermal tuning units and a sophisticated control algorithm \cite{27}.

At present, most on-chip polarization controllers still rely on blind search-based methods, similar to those used in fiber devices. In contrast, few studies have explored optical-computing approaches for DPCs on photonics chips. In 2025, Aloe Semiconductor designed an optical computing method using three phase shifters to achieve polarization locking \cite{6,7}. The feedback process is based on measuring, analyzing, and rotating the Stokes vectors. To distinguish this approach from blind search-based methods, it is referred to APC method hereafter.

The APC method introduces optical computing into polarization locking, enabling any input polarization state to be transformed into the desired polarization state within a single control loop. In particular, it can overcome the limitations imposed by the slow response of thermal phase shifters used for polarization locking in silicon photonics chips, which offer the advantages of low cost and complementary metal-oxide-semiconductor (CMOS) compatibility \cite{34}. However, the previously reported APC method did not sufficiently explore the method of calibration and endless control. In particular, the last MZI used in Stokes-vector measurement did not include a phase shifter. Consequently, the phase difference within this MZI cannot be compensated, thereby reducing the accuracy of polarization control.

In this work, we design and simulate an APC method with four phase shifters on a photonics chip. The basic structures and operations of APC are outlined. The calibration method and the endless control method are explored and simulated in detail. The phase difference involved in measuring Stokes vectors can be compensated with the fourth phase shifter in the last or third MZI structure, and the output extinction ratio induced by the fourth phase difference is analyzed. In addition, the adding of fourth phase shifter can implement the rotation operation around all three axes of Poincaré sphere, opening the door to exploring more APC methods through optical computing.

\section{Polarization-state representations and basic operations}
\subsection{Jones vectors and Stokes vectors}
It is well known that a pure polarization state can be represented by the normalized Jones vector as follows:
\begin{equation}
J(\delta,\alpha)=
\begin{bmatrix}
E_x\\
E_y
\end{bmatrix}
=
\begin{bmatrix}
\cos\alpha\\
e^{i\delta}\sin\alpha
\end{bmatrix},
\tag{1}
\end{equation}
where the phase $\alpha\in[0,\pi/2]$ determines the amplitudes of the two polarization components 
$|E_x|=\cos\alpha$ and $|E_y|=\sin\alpha$, and the phase 
$\delta\in[0,2\pi)$ represents the relative phase between the two polarization components. The Jones vector is related to the normalized Stokes vector
$\mathbf{S}(\delta,\alpha)=[S_0,S_1,S_2,S_3]^{T}$:
\begin{equation}
\begin{aligned}
S_0 &= |E_x|^2+|E_y|^2=1,\\
S_1 &= |E_x|^2-|E_y|^2=S_0\cos2\alpha,\\
S_2 &= 2|E_x||E_y|\cos\delta=S_0\sin2\alpha\cos\delta,\\
S_3 &= 2|E_x||E_y|\sin\delta=S_0\sin2\alpha\sin\delta .
\end{aligned}
\tag{2}
\label{eq:2}
\end{equation}

\begin{figure}[!ht] 
	\centering\includegraphics[width=8.8cm]{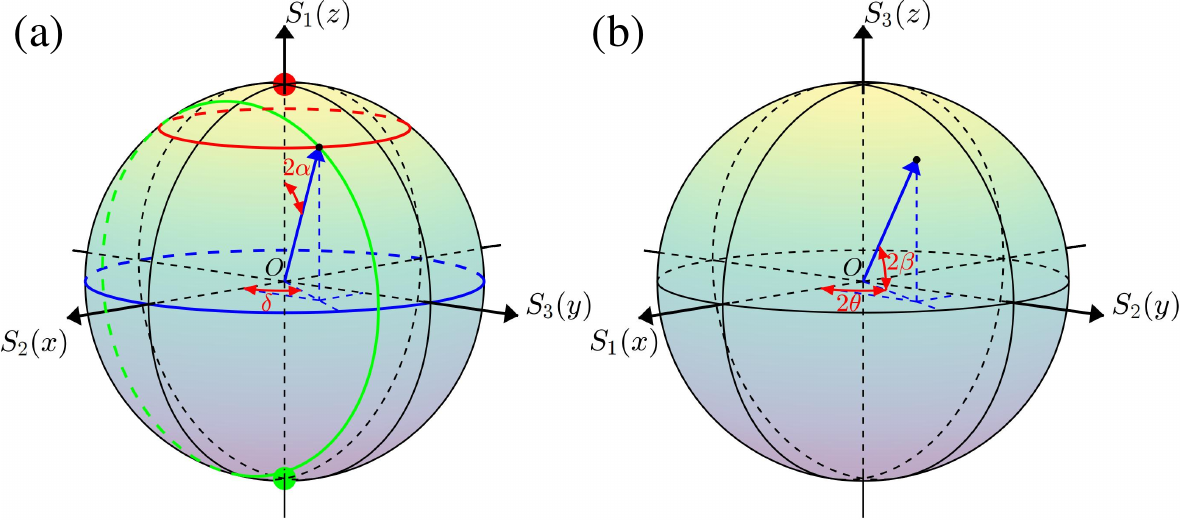}\\
	\caption{\label{1} Two types of Poincaré spheres. (a) The Poincaré sphere with phases $\delta$ and $2\alpha$; (b) The Poincaré sphere with phases $2\theta$ and $2\beta$.}
\end{figure}

The Stokes vector can be represented as a blue arrow in the Poincaré sphere shown in Fig.~\ref{1}(a). The $S_1$ axis points upward in the $z$ direction, and the $S_2$ and $S_3$ axes are in the $x$ and $y$ directions, respectively. To describe the vectors and sphere more vividly, the point where the positive semi $S_1$ axis intersects the spherical surface is denoted as the north pole(red point), and the point where the negative semi $S_1$ axis intersects the spherical surface is denoted as the south pole(green point). The phase $2\alpha$ between the Stokes vector and the $S_1$ axis is called the latitude phase. The plane formed by the $S_2$ and $S_3$ axes is the equatorial plane, and the blue line where this plane intersects the sphere is denoted as the equator. The phase $\delta$ between the projection of the Stokes vector onto the equatorial plane and the $S_2$ axis is called the longitude phase. For general polarization control, the polarization state will be locked to the north pole or near the north pole corresponding to a latitude phase $2\alpha$ around 0 (linear polarization state). 

The Stokes vector provides a complete representation of any polarization state. Specifically, a pure polarization state corresponds to a vector whose endpoint resides on the surface of the Poincaré sphere, whereas a mixed polarization state is denoted by an endpoint inside the sphere. In our manuscript, we focus on the pure polarization state. 


It should be noted that there is another representation of the Poincaré sphere as shown in Fig.~\ref{1}(b), which differs from the representation method used in our manuscript. The Stokes vector can be represented by 
\begin{equation}
\begin{aligned}
S_0 &= 1,\\
S_1 &=S_0\cos2\theta\cos2\beta,\\
S_2 &=S_0\sin2\theta\cos2\beta,\\
S_3 &=S_0\sin2\beta,
\end{aligned}
\tag{3}
\end{equation}
where the phase $\theta$ denotes the orientation phase or azimuth phase, and $\beta$ denotes the ellipticity phase. Further differences between the two types of representations can be found in literature \cite{35}. 

\subsection{Typical operations used in DPC in fiber and block optics}
There are many devices and corresponding operations for DPC in fiber and block optics \cite{36,37,38,39}. Here, we focus on operations that delay the phase. The typical devices are 0° and 45° wave plates or equivalent structures. Their effects on the polarization state can be represented by Jones matrices as 
\begin{equation}
\begin{aligned}
J_0=\begin{bmatrix} 1 & 0 \\ 0 & e^{i\delta} \end{bmatrix}, \quad 
J_{45} = \frac{1}{2} \begin{bmatrix} 1 + e^{i\delta} & 1 - e^{i\delta} \\ 
1 - e^{i\delta} & 1 + e^{i\delta} \end{bmatrix}.
\end{aligned}
\tag{4}
\end{equation}
The corresponding Mueller matrices can be derived using Eqs.~\ref{eq:5} and~\ref{eq:6} \cite{40}
\begin{equation}
\begin{aligned}
M_0 &= A(J_0 \otimes J_0^*) A^\dagger \\
&= \begin{bmatrix}
1 & 0 & 0 & 0 \\
0 & 1 & 0 & 0 \\
0 & 0 & \cos\theta & \sin\theta \\
0 & 0 & -\sin\theta & \cos\theta
\end{bmatrix},
\end{aligned}
\tag{5}
\label{eq:5}
\end{equation}

\begin{equation}
\begin{aligned}
M_{45} &= A(J_{45} \otimes J_{45}^*) A^\dagger \\
&= \begin{bmatrix}
1 & 0 & 0 & 0 \\
0 & \cos\theta & 0 & -\sin\theta \\
0 & 0 & 1 & 0 \\
0 & \sin\theta & 0 & \cos\theta
\end{bmatrix},
\end{aligned}
\tag{6}
\label{eq:6}
\end{equation}
where the matrix $A$ is
\begin{equation}
A = \frac{1}{\sqrt{2}}
\begin{bmatrix}
1 & 0 & 0 & 1 \\
1 & 0 & 0 & -1 \\
0 & 1 & 1 & 0 \\
0 & i & -i & 0
\end{bmatrix}.
\tag{7}
\end{equation}

When \(\theta > 0\), the matrix \(M_0\) rotates the Stokes vector clockwise around the \(S_1\) axis, as shown by the red trajectory in Fig.~\ref{1}(a). The matrix \(M_{45}\) rotates the Stokes vector counterclockwise around the \(S_2\) axis, as shown by the green trajectory. With these two basic operations, any pure polarization state can be rotated to any desired pure polarization state.

\section{Structures and operations on a photonics chip}
\subsection{Basic structures and Rotation around the $S_1$ axis} 
\begin{figure}[!ht] 
	\centering\includegraphics[width=8cm]{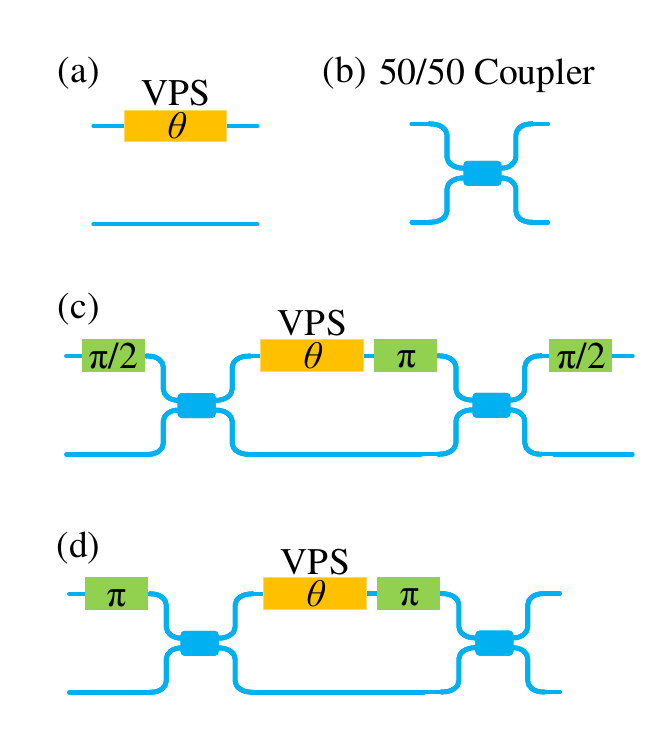}\\
	\caption{\label{2}The basic structures and rotation structures on photonics chip (a) Parallel waveguides and structure for rotation around the $S_1$  axis; (b) 2 × 2 50/50 coupler; (c) Optical waveguide structure for rotation around the $S_2$ axis; (d) Optical waveguide structure for rotation around the $S_3$ axis. VPS: Variable phase shifter.}
\end{figure}
In optical waveguides, the two orthogonal polarization components propagate in two separate waveguides, which are usually arranged in parallel. The two basic operating structures are a phase shifter in parallel waveguides (Fig.~\ref{2}(a)) and a 2×2 50/50 coupler (Fig.~\ref{2}(b)). Their operations can be represented by Jones matrices $J_{PS}$ and $J_{CO}$:
\begin{equation}
\begin{aligned}
J_{PS} = J_{S1} &= 
\begin{bmatrix}
e^{-i\theta} & 0 \\
0 & 1
\end{bmatrix}
=
\begin{bmatrix}
1 & 0 \\
0 & e^{i\theta}
\end{bmatrix}
e^{-i\theta},
\end{aligned}
\tag{8}
\label{eq:8}
\end{equation}

\begin{equation}
\begin{aligned}
J_{CO} &= \frac{1}{\sqrt{2}}
\begin{bmatrix}
1 & -i \\
-i & 1
\end{bmatrix}.
\end{aligned}
\tag{9}
\label{eq:9}
\end{equation}

Because a common phase does not affect the polarization state, the two forms of matrix $J_{PS}$ are equivalent. To present the modes concisely, propagation loss in waveguide and coupling-ratio deviations are not considered. 

The Mueller matrices of above two basic structures are
\begin{equation}
\begin{aligned}
M_{PS} = M_{S1} &= 
\begin{bmatrix}
1 & 0 & 0 & 0 \\
0 & 1 & 0 & 0 \\
0 & 0 & \cos\theta & \sin\theta \\
0 & 0 & -\sin\theta & \cos\theta
\end{bmatrix},
\end{aligned}
\tag{10}
\label{eq:10}
\end{equation}

\begin{equation}
\begin{aligned}
M_{CO} &= 
\begin{bmatrix}
1 & 0 & 0 & 0 \\
0 & 0 & 0 & -1 \\
0 & 0 & 1 & 0 \\
0 & 1 & 0 & 0
\end{bmatrix}.
\end{aligned}
\tag{11}
\label{eq:11}
\end{equation}

Obviously, the phase shifter in parallel waveguides can perform the operation of rotation around the \(S_1\) axis. The other two types of structures that rotate around axes \(S_2\) or \(S_3\) can be realized through composite structures of basic operating structures, which will be discussed below.  

\subsection{Rotation around the $S_2$ axis} 

Figure~\ref{2}(c) shows the structure of rotation around the \(S_2\) axis in optical waveguides. It is a typical MZI with one variable phase shifter and three additional fixed phase shifters $\pi/2, \pi$, and $\pi/2$. All phase shifters are located in the upper waveguides. The Jones matrix of the structure is 
\begin{equation}
J_{S2} = -\frac{1}{2} \begin{bmatrix}
1 + e^{i\theta} & 1 - e^{i\theta} \\
1 - e^{i\theta} & 1 + e^{i\theta}
\end{bmatrix}.
\tag{12}
\end{equation}
The Mueller matrix of the structure can be expressed as
\begin{equation}
M_{S2} =
\begin{bmatrix}
1 & 0 & 0 & 0 \\
0 & \cos\theta & 0 & -\sin\theta \\
0 & 0 & 1 & 0 \\
0 & \sin\theta & 0 & \cos\theta
\end{bmatrix}.
\tag{13}
\end{equation}
When \( \theta > 0 \), the matrix \( M_{S2} \) causes the Stokes vector to rotate anticlockwise around the \(S_2\) axis. 
\subsection{Rotation around the $S_3$ axis} 
Figure~\ref{2}(d) shows the rotation structure around the \(S_3\) axis in optical waveguides. It is a typical MZI with one variable phase shifter and two additional constant phase shifters $\pi$.

All phase shifters are located in the upper waveguides. The Jones matrix of the structure is 
\begin{align}
J_{S3} &= J_{CO} \cdot J_{PS}(\pi) \cdot J_{S1} \cdot J_{CO} \cdot J_{PS}(\pi) \nonumber \\
&= \frac{1}{2} \begin{bmatrix} -1 - e^{i\theta} & ie^{i\theta} - i \\ ie^{i\theta} - i & 1 - e^{i\theta} \end{bmatrix}.
\tag{14}
\end{align}

The Muller matrix $M_{S3} $ is derived as follows: 
\begin{equation}
M_{S3} =
\begin{bmatrix}
1 & 0 & 0 & 0 \\
0 & \cos\theta & \sin\theta & 0 \\
0 & -\sin\theta & \cos\theta & 0 \\
0 & 0 & 0 & 1
\end{bmatrix}.
\tag{15}
\end{equation}
When \( \theta > 0 \), the matrix \( M_{S3} \) causes the Stokes vector to rotate clockwise around the \(S_3\) axis. 

\subsection{Structure for measuring the Stokes vector} 
\begin{figure}[!ht] 
	\centering\includegraphics[width=8.8cm]{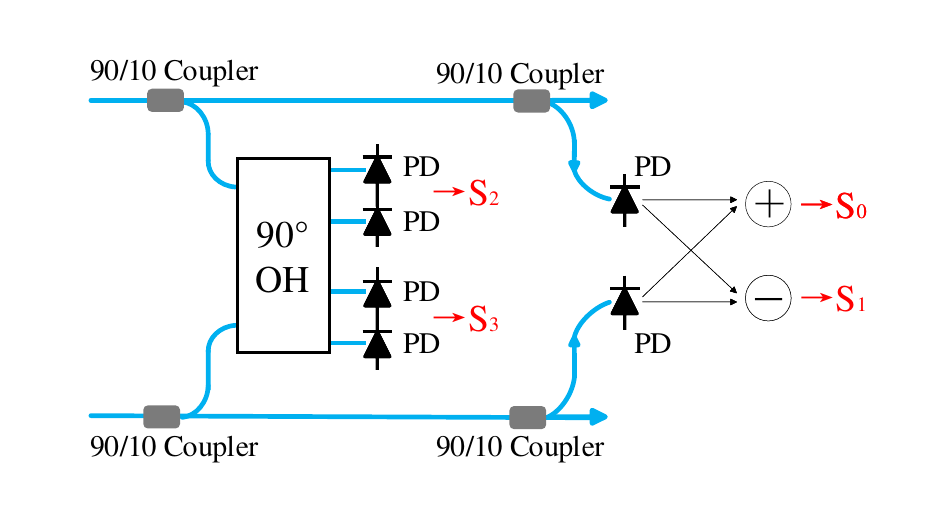}\\
	\caption{\label{3} Structure for measuring the Stokes vectors. OH: optical hybrid, PD: photodiodes.}
\end{figure}

Figure~\ref{3} shows the structure used to measure the Stokes vectors $S$. Four couplers separate the beams into two pairs. One pair of beams is guided into one 90° optical hybrid with two pairs of photodiodes, which are mainly used to measure the elements $S_2$ and $S_3$ of the Stokes vectors $S$. The splitting ratio is $R_1$. The other pair of beams is guided into two separate photodiodes to measure the elements $S_0$ and $S_1$  of the Stokes vectors $S$. The splitting ratio is $R_2$.

The electric fields in the upper and lower waveguides are $E_x$ and  $E_y$, and their intensities are \( I_x = |E_x|^2 \) and \( I_y = |E_y|^2 \), respectively. The components $S_0$ and $S_1$ can be determined by 
\begin{equation}
\begin{aligned}
S_0 &= \frac{(1-R_1)R_2 I_x + (1-R_1)R_2 I_y}{(1-R_1)R_2} = I_x + I_y ,\\
S_1 &= \frac{(1-R_1)R_2 I_x - (1-R_1)R_2 I_y}{(1-R_1)R_2} = I_x - I_y.
\end{aligned}
\tag{16}
\end{equation}

Using the 90° optical hybrid, the components $S_2$ and $S_3$  can be determined by
\begin{equation}
\begin{aligned}
S_2 &= \frac{\sqrt{R_1} |E_x| \sqrt{R_1} |E_y| \cdot \cos \delta}{R_1 / 2} = 2 \cdot |E_x||E_y| \cos \delta,\\
S_3& = \frac{\sqrt{R_1} |E_x| \sqrt{R_1} |E_y| \cdot \sin \delta}{R_1 / 2} = 2 \cdot |E_x||E_y| \sin \delta.
\end{aligned}
\tag{17}
\end{equation}

In the above mentioned four equations, The terms \( (1-R_1) R_2 \) and \(R_1/2\) are used to calibrate the measurement results. By normalizing them with \( S_0 \) and setting $|E_x| / \sqrt{S_0} = \cos \alpha$, we can obtain the Stokes vectors as in Eq.~\ref{eq:2}.

The Stokes vector \( S_{\text{out}} \) differs from the measured $S_m$ owing to an additional unknown longitudinal phase. When the polarization state $S_m$ is locked to the north pole, the output intensity remains in the upper waveguide and is not affected by the additional longitudinal phase, or \( S_{\text{out}} \) is also at the north pole.  
\begin{figure*}[!ht] 
\centering\includegraphics[width=16cm]{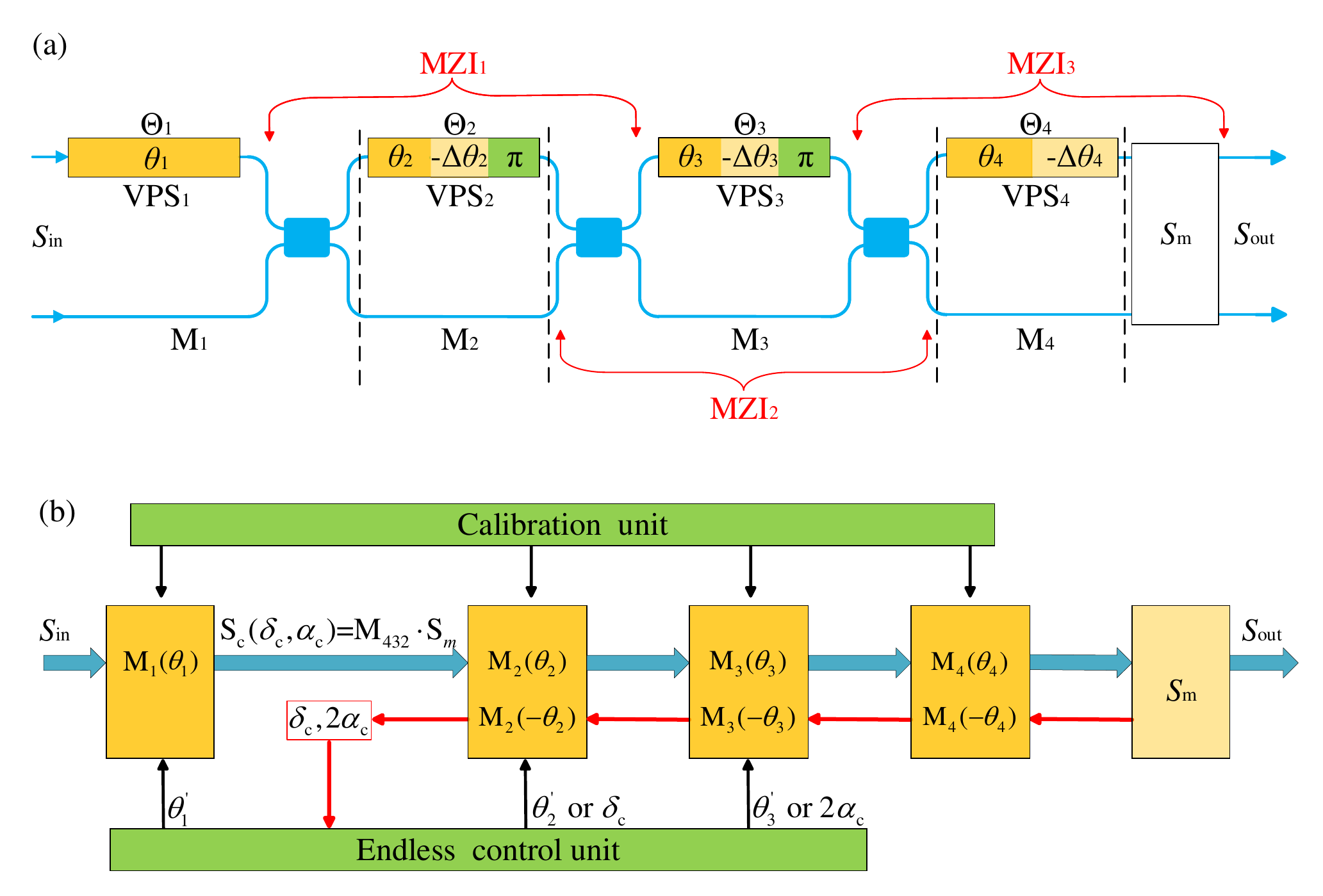}\\
 \caption{\label{4} Structure (a) and  relationship of various functional units (b) in APC.}
\end{figure*}

\section{Structure and units for APC} 
\subsection{Introduction of structure and units for APC}

Based on the basic rotation and measurement units in Section 3, we design the structure for APC, as shown in Fig.~\ref{4}(a). It is mainly composed of four control units with four phase shifts \( \Theta_1 \), \( \Theta_2 \), \( \Theta_3 \), \( \Theta_4 \) and one measurement unit. The function of each control unit can be described by Mueller matrices \( M_1(\theta_1) \), \( M_2(\theta_2) \), \( M_3(\theta_3) \), and \( M_4(\theta_4) \). Here, \( \theta_1, \theta_2, \theta_3, \) and \(\theta_4 \) are the control phases of the control units. These units transform the input Stokes vectors $S_{\text{in}}$ into Stokes vectors $S_{\text{out}}$. The Stokes vector after the first control unit is $S_c(\delta_c, \alpha_c)$, and the measured Stokes vector is denoted as $S_m$. Their relationships are plotted in Fig.~\ref{4}(b). The calibration unit will calibrate the phase differences $\Delta\theta_2, \Delta\theta_3$, and $\Delta\theta_4$ of three cascaded MZI structures, while the endless control unit will keep the control process running continuously, and the output intensities will not be affected by the limited control phase.       

The first control unit is the endless-control auxiliary unit, composed of a phase shifter $\Theta_1$ with control phase $\theta_1$ in the upper waveguide and a 50/50 coupler. Its function is explained in detail in the the subsection on the endless polarization control process.

The second unit is used to rotate the Stokes vector \( S_c \) around the \( S_1 \)  axis. Thus, the phase \( \delta_c \) can be rotated to \( \delta'_c = \delta_c - \theta_2 \). When \( \theta_2 = \delta_c \), the phase \( \delta'_c \) becomes zero. The phase shifter $\Theta_2$ in this unit not only provides the control phase \( \theta_2 \), but also provides the compensation phase \( -\Delta\theta_2 \) and the constant phase shift \(\pi\). The phase shifter $\Theta_2$ in the second unit provides the constant phase \( \pi \), which is then used together with $\Theta_3$ in the third unit to rotate \( \alpha \).

The third control unit is used to rotate the Stokes vector  \( S_c \) around the \( S_3 \) axis. When the phase \( \delta_c \) is rotated to zero, the phase \( \alpha_c \) can be accurately rotated to \( \alpha'_c = \alpha_c - \theta_3 / 2\) by tuning the control phase $\theta_3$. When \( \theta_3 = 2\alpha_c \), the phase \( \alpha'_c \) will be zero. The phase shifter $\Theta_3$ in this unit not only provides the control phase $\theta_3$, but also provides the compensation phase \( -\Delta\theta_3 \) and the constant phase shift $\pi$. Here, rotation around the axis \( S_3 \) is utilized. If rotation around \( S_2 \) is used, the related constant phase shift in the second and fourth shifters should be \( \pi/ 2 \).

The fourth control unit is usually used to compensate for phase differences in the measurement of the Stokes vector. In our control process, rotation around the $S_3$ axis is used. The fourth phase shifter $\Theta_4$ need not provide an additional phase shift for the third unit. The fourth phase shifter only provides the compensation phase \( -\Delta\theta_4 \). If rotation around the $S_2$ axis is used, an additional phase shift \( \pi / 2 \) should be provided by the fourth unit for the third unit’s function. However, when \(\Delta\theta_4 \) is positive, the thermal phase shifter cannot provide a negative phase to compensate for the phase difference. Thus, a control phase greater than \(\Delta\theta_4 \) should be added when a thermal phase shifter is used, such that \(\theta_4 = \pi/2 \) can be selected. The Stokes vector after the third unit will be at the north pole. The fourth unit rotates the Stokes vectors around the $S_1$ axis near the north pole with a phase \(\theta_4 \), which will not affect the output intensity in the upper or lower waveguide. 

The measurement unit is used to measure the Stokes vectors. In the control process, the Stokes vectors are rotated to the north pole. We should use the inverse functions $M(-\theta_2), \ M(-\theta_3) \ \text{and} \ M(-\theta_4)$ in Eq.\ref{eq:18} to derive the Stokes vectors $S_c$ before the second control unit
\begin{equation}
S_c(\delta_c, \alpha_c) = M_2(-\theta_2) \cdot M_3(-\theta_3) \cdot M_4(-\theta_4) \cdot S_m.
\tag{18}
\label{eq:18}
\end{equation}

Based on the known Stokes vector \( S_c(\delta_c, \alpha_c) \), the second and third control phases are adjusted to \( \theta_2 = \delta_c, \ \text{and} \
\theta_3 = 2\alpha_c \). At this stage, the Stokes vector \( S_c(\delta_c, \alpha_c) \) is rotated to the north pole.

\subsection{The calibration process}

\begin{figure}[!ht] 
	\centering\includegraphics[width=9cm]{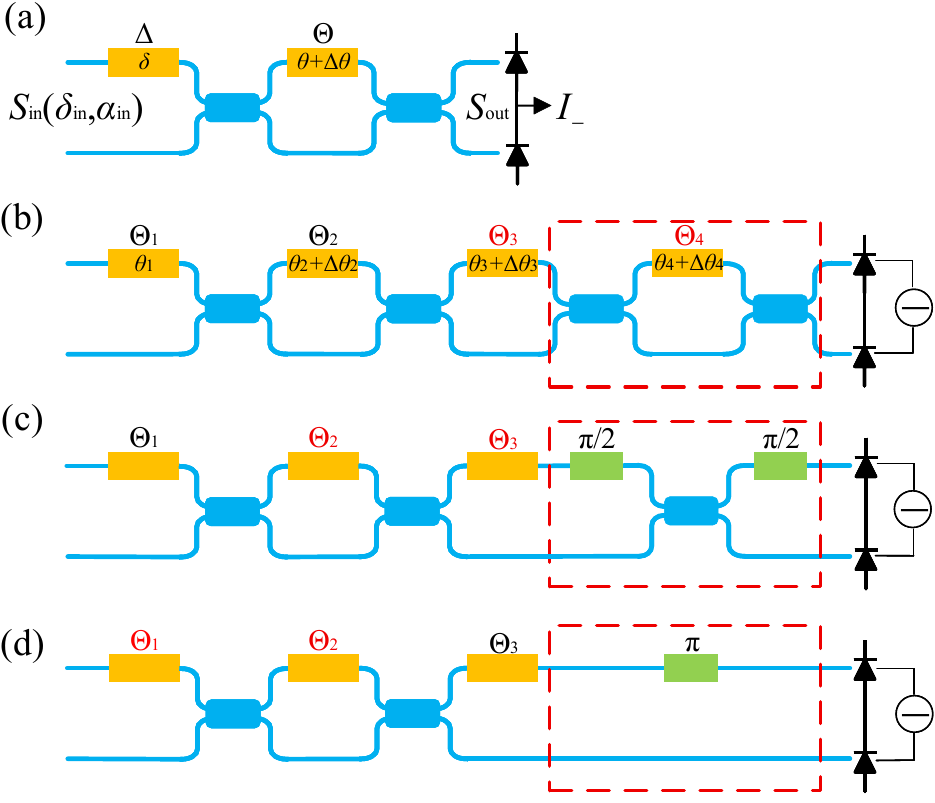}\\
	\caption{\label{5}Various structures for calibration. (a) Basic structure of the pairwise scanning method; (b) Equivalent cascaded MZI structures; (c) Equivalent cascaded MZI structures when \(\theta_4+\Delta\theta_4=\pi/2\); (d) Equivalent cascaded MZI strugtures when \(\theta_4+\Delta\theta_4=\pi\).}
\end{figure}

The on-chip optical structures used for APC are cascaded MZI structures. To calibrate them, the pairwise scan method was refined and applied. Figure~\ref{5}(a) shows the basic structure of the pairwise scan method. There is one phase shifter \( \Delta \) outside the MZI structure and one phase shifter \( \Theta \) inside the MZI structure. Their control phases are denoted as \( \delta \) and \( \theta \), and the phase difference inside the MZI structure is denoted as \( \Delta\theta \). The input polarization state is denoted as \(S_{in}\). The output Stokes vectors can be derived as
\begin{equation}
\resizebox{0.88\linewidth}{!}{$\displaystyle
\begin{aligned}
S_{\mathrm{out}} &= M_{\mathrm{CO}} \cdot M_{\mathrm{S1}}(\theta) \cdot M_{\mathrm{CO}} \cdot M_{\mathrm{S1}}(\delta) \cdot S_{\mathrm{in}} \\
&= \begin{bmatrix}
1 \\
c_1 \sin(\theta+\Delta\theta)\cos(\delta_{\mathrm{in}}-\delta)-c_2\cos(\theta+\Delta\theta) \\
c_2 \sin(\theta+\Delta\theta) + c_1 \cos(\theta+\Delta\theta)\cos(\delta_{\mathrm{in}}-\delta) \\
-c_1 \sin(\delta_{\mathrm{in}}-\delta)
\end{bmatrix}.
\end{aligned}
$}
\tag{19}
\label{eq:19}
\end{equation}

The difference in the intensities of the two outputs is
\begin{equation}
\begin{split}
I_- &= S_1 \\
&= c_{1}\sin(\theta + \Delta\theta) \cos(\delta_{in} - \delta) -c_{2} \cos(\theta + \Delta\theta).
\end{split}
\tag{20}
\label{eq:20}
\end{equation}
where \( c_{1}=sin2\alpha_{in}\), \( c_{2}=cos2\alpha_{in}\), and \(c_1^2 + c_2^2 = 1\). Since the input Stokes state remains stable during calibration, the phases  \( \alpha_{in}\) and \( \delta_{in} \) are constant. In pairwise scanning, the phase \(\theta \) is in the outer loop with a scanning range \([0, \pi]\), and the phase \( \delta \) is in the inner loop with a scanning range \([0, 2\pi]\). For each value of \( \theta_i \), we can obtain a peak-to-peak value \(I_{PP}\) of the output intensities \(I_-\) as follows.
\begin{equation}
I_{PP} = \left|2 c_{1} \cdot \sin(\theta_i + \Delta\theta) \right|.
\tag{21}
\label{eq:21}
\end{equation}

The minimum value $I_{\text{PPmin}}$ corresponds to $\theta_{i_{P0}} + \Delta\theta = 0$ or $\theta_{i_{P\pi}} + \Delta\theta = \pi$. The power applied to phase shifter $\theta$ is denoted by $P_{\theta\text{min}}$. The maximum value $I_{\text{PPmax}}$ corresponds to $\theta_{i_{P\pi/2}} + \Delta\theta = \pi/2$. The power applied to the phase shifter $\theta$ is denoted as $P_{\theta\text{max}}$. The following rules can be used to calibrate phase shifter $\theta$:

When $P_{\theta\text{min}} < P_{\theta\text{max}}$,
\begin{equation}
\begin{split}
\theta_{i_{P0}} + \Delta\theta = 0
&\implies \Delta\theta = -\theta_{i_{P0}} \\
&\implies \Delta\theta = -P_{\text{min}} \cdot k_\theta.
\end{split}
\tag{22}
\label{eq:22}
\end{equation}

When $P_{\theta\text{min}} > P_{\theta\text{max}}$,
\begin{equation}
\begin{split}
\theta_{i_{P\pi}} + \Delta\theta = \pi
&\implies \Delta\theta = \pi - \theta_{i_{P\pi}} \\
&\implies \Delta\theta = \pi - P_{\text{min}} \cdot k_\theta.
\end{split}
\tag{23}
\label{eq:23}
\end{equation}
\begin{figure}[!ht] 
	\centering\includegraphics[width=8.8cm]{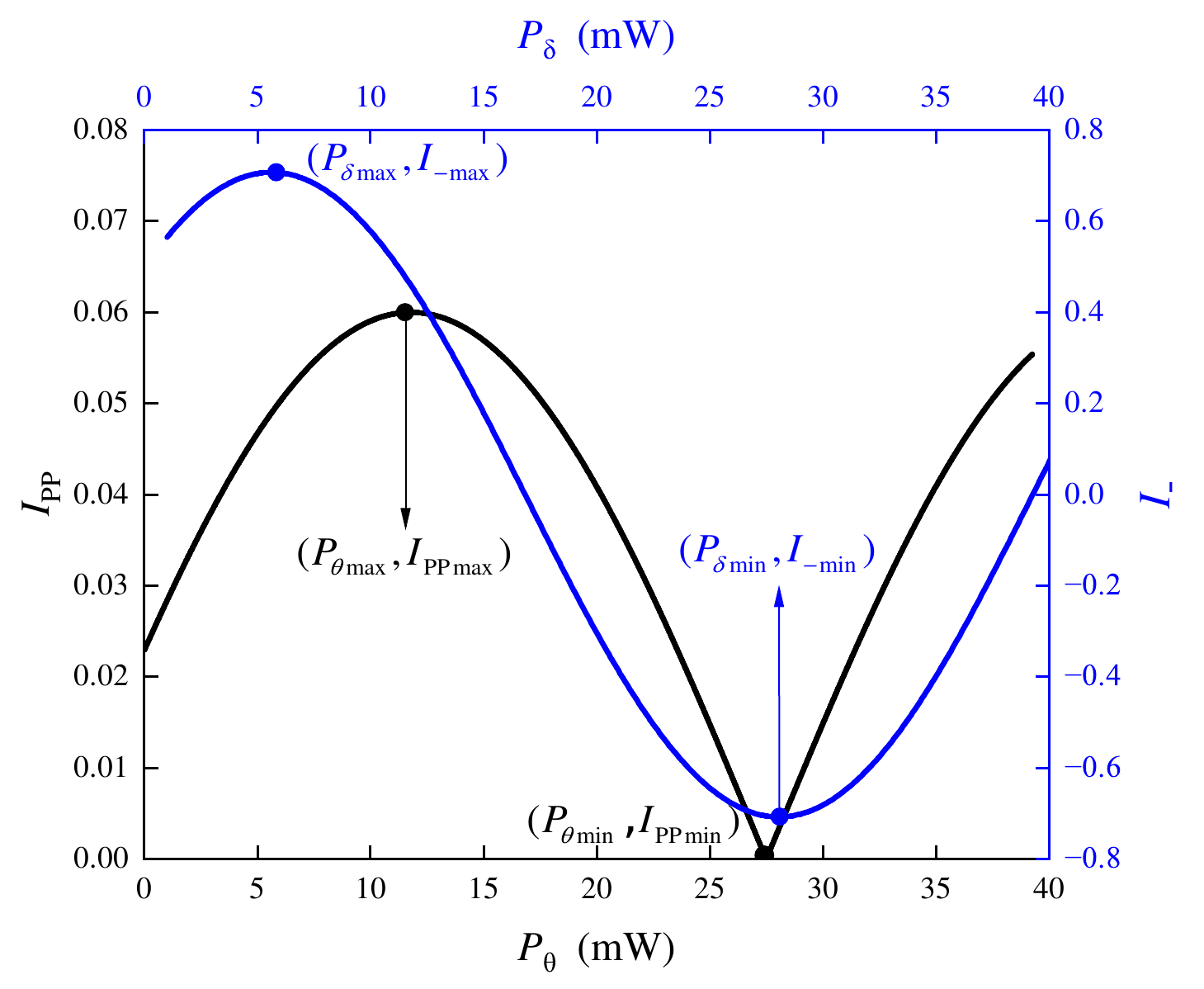}\\
	\caption{\label{6} Simulation results of the pairwise scan method for calibrating phase shifters $\Theta$ and $\Delta$.}
\end{figure}

In both cases,
\begin{equation}
\begin{split}
&|P_{\theta\text{max}} - P_{\theta\text{min}}\bigr| \cdot k_\theta = \pi/2 \\
&\implies k_\theta = (\pi/2)\Big/\bigl|P_{\theta\text{max}} - P_{\theta\text{min}}|.
\end{split}
\tag{24}
\label{eq:24}
\end{equation}

After determining the phase difference $\Delta\theta$ and slope $k_\theta$ of the phase shifter $\Theta$, set its value to \( \theta_{i_P\pi/2} + \Delta \theta = \pi/2 \). The output intensity \( I_- \) is
\begin{equation}
I_- =c_{1} \cos(\delta_{in} - \delta).
\tag{25}
\label{eq:25}
\end{equation}
Then scan the phase shifter \( \Delta\) in the range \([0, 2.5\pi]\). Note that the power \( P_{\delta_{\text{max}}} \) corresponds to the maximum output value \( I_{-\text{max}} \), and the power \( P_{\delta_{\text{min}}} \) corresponds to the minimum output value \( I_{-\text{min}} \). The slope \( k_\delta \) of the phase shifter \(\Delta \) can be determined by
\begin{equation}
\left| P_{\delta_{\text{max}}} - P_{\delta_{\text{min}}} \right| \cdot k_\delta = \pi
\tag{26}
\label{eq:26}
\end{equation}

In Fig~\ref{6}, a simulation was performed to evaluate the calibration method. In the simulation, the phases of the input Stokes vector were set to \( \delta_{in} = \pi/4 \), and \( \alpha_{in} = \pi/8 \). The phase difference was set to \(\Delta\theta = \pi/8\), and the slopes \( k_{\delta} \) and \( k_{\theta} \) were set to 0.14 mW/rad. The black curve in Fig~\ref{6} shows the simulation results for calibrating the phase shifter \(\Theta \). The evaluated parameters are listed in Table~\ref{tab:1}. The accuracy of the evaluated parameters depends on the accuracy of the scanning step \cite{41}. The smaller the scanning step size, the more accurate the evaluated parameters. However, more time is required. In the simulation, the scanning step was set to 0.01 rad.

\begin{table}[htbp]
    \centering
    \caption{The evaluated parameters in calibrating the phase shifter \(\Theta\).}
    \label{tab:1}
    \small
    \begin{tabular}{cccccc}
        \hline
        \( P_{\theta\max} \) & \( I_{PP\theta\max} \) & \( P_{\theta\min} \) & \( I_{PP\theta\min} \) & \( k_{\theta} \) & \( \Delta\theta \) \\
        (mW) & & (mW) & & (rad/mW) & \\
        \hline
        8.42    & 1.4142   & 19.62    & 0.0030  & 0.1402 & 0.3948\\
        \hline
    \end{tabular}
\end{table}

The blue curve in Fig~\ref{6} shows the simulation results for calibrating the phase shifter \(\Delta \). The evaluated parameters are listed in Table~\ref{tab:2}. In the simulation, the scanning step is set to 0.01 rad.
\begin{table}[htbp]
    \centering
    \caption{The evaluated parameters in calibrating the phase shifter \(\Delta\).}
    \label{tab:2}
    \small
    \begin{tabular}{ccccc}
        \hline
        \( P_{\delta_{\text{max}}} \) (mW) & \( I_{-\text{max}} \) & \( P_{\delta_{\text{min}}} \) (mW) & \( I_{-\text{min}} \) & \( k_{\delta} \) (rad/mW) \\
        \hline
        5.62    & 0.7071    & 28.02    & 0.7071    & 0.1402 \\
        \hline
    \end{tabular}
\end{table}

Based on the above calibration method, we can calibrate the pair of phase shifters \(\Theta_3\) and \(\Theta_4\)  firstly in experiment as shown in Fig.~\ref{5}(b). After calibrating the first pair, we set the phase of fourth phase shifter to \(\theta_4+\Delta\theta_4= \pi/2\). The third MZI structure will be equivalent to a 2×2 MMI structure with two additional \(\pi/2\) phase shifter, as shown in Fig.~\ref{5}(c) \cite{41}. Then the pair of phase shifters \(\Theta_2\) and \(\Theta_3\) can be calibrated. After calibrating the second pair, we set the fourth phase shifter to \(\theta_4+\Delta\theta_4= \pi\). The third MZI structure will be equivalent to a parallel structure with one additional \(\pi\) phase shifter as shown in Fig.~\ref{5}(d). Then the pair of phase shifters \(\Theta_1\) and \(\Theta_2\) can be calibrated.

\subsection{The endless polarization control process}

\begin{figure}[!ht] 
	\centering\includegraphics[width=8.8cm]{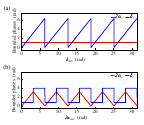}\\
	\caption{\label{7} Phase jump of the bounded longitude phase $\delta_c$ caused by the variation of actual phases $\alpha_{act}$ (a) and $2\alpha_{act}$ (b).}
\end{figure}

To present the Stokes vectors on the Poincaré sphere and calculate the Stokes parameters conveniently, the scale of the longitude phase \( \delta_c \) is bounded in the range \([0, 2\pi)\), and the scale of the latitude phase \( 2\alpha_c \) is bounded in the range \([0, \pi]\). In actual situations, the phases of the polarization states \( \delta_{\text{act}} \) and \( 2\alpha_{\text{act}} \) are not bounded by the above ranges; the actual range is \((-\infty, +\infty)\). The difference between the actual and bounded phases will lead to phase jumps. For example, when the actual longitude \( \delta_{\text{act}}\) changes from \( 2n\pi-\varepsilon\) to \( 2n\pi + \varepsilon\) continuously, the bounded longitude phase \(\delta_c\) jumps from \(2\pi- \varepsilon\) to \(\varepsilon\), as shown in Fig.~\ref{7}(a). Here, \(\varepsilon\) represents a small phase. The actual latitude phase \( 2\alpha_{\text{act}} \) remains constant, and the bounded latitude phase \( 2\alpha_c \) also remains constant. When the actual latitude phase \( 2\alpha_{\text{act}} \) changes from \( n\pi - \varepsilon\) to \( n\pi + \varepsilon \) continuously, the bounded longitude phase \( 2\alpha_c \) changes continuously from \( \pi -\varepsilon\) to \( \pi - \varepsilon \) or from \( 0 + \varepsilon \) to \( 0 + \varepsilon \), as shown in Fig.~\ref{7}(b), and there is no jump. Although the actual longitude phase \( \delta_{\text{act}} \) remains constant, the bounded longitude phase jumps from \( \delta_c \) to \( \delta_c + \pi \) or \( \delta_c + \pi \) to \( \delta_c \).

\begin{figure}[!ht] 
	\centering\includegraphics[width=8.8cm]{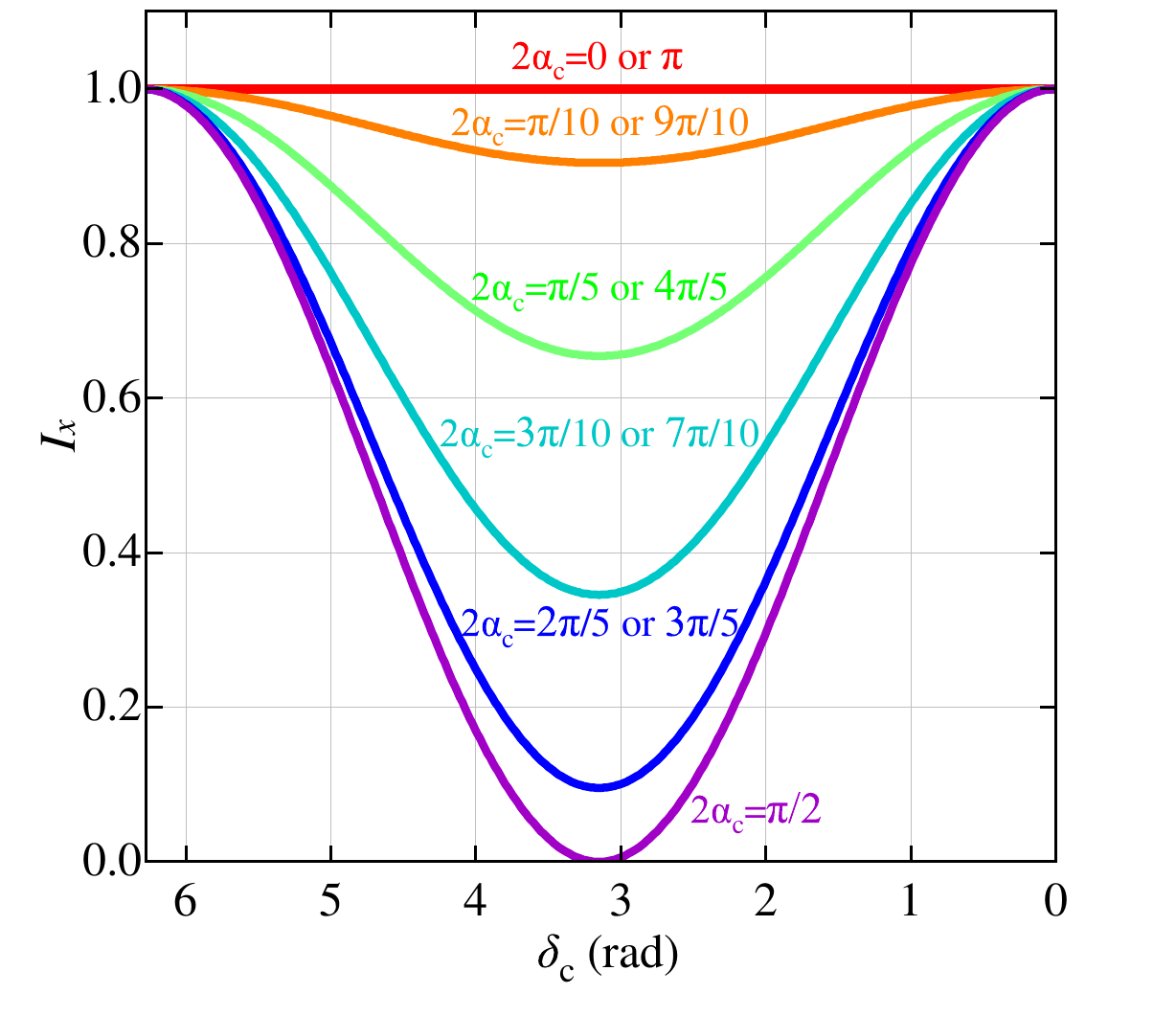}\\
	\caption{\label{8}Normalized output intensity versus the longitude phase \( \delta_c\) at different latitude phases.}
\end{figure}

In actual physical situation, these jumps require a duration, during which the jumping value changes step by step. This process will lead to intensity disturbances in polarization control. To observe the disturbance in detail when the bounded longitude phase \( \delta_c \) jumps from \( 2\pi - \varepsilon\) to \( \varepsilon\), we divide the jump into 360 steps. Figure~\ref{8} shows the intensity changes versus the longitude phase \( \delta_c \)  at different latitude phases \( 2\alpha_c \). When the latitude phase \( 2\alpha_c \) is 0 or \(\pi\), corresponding to the north pole or the south pole, the normalized output intensity does not change with the longitude value, and the jumping introduces no disturbance, as shown by the red line in Fig.~\ref{8}. For other latitude phases, the normalized output intensity varies from the maximum value to the minimum value in the range \((\pi, 2\pi)\) of longitude values \(\delta_c\) and from the minimum value to the maximum value in the range \([0,\pi]\). When the latitude phase \(2\alpha_c\) is in the range \([0, \pi/2]\), the minimum value decreases as the latitude phase increases. When the latitude phase \(2\alpha_c= \pi/2\), the minimum value can decrease to zero, which results in the maximum disturbance. When the latitude \(2\alpha_c\) is in the range \([\pi/2, \pi]\), the minimum value increases as the latitude phase increases. Here, we analyze the case in which \(\delta_c\) jumps form \(2\pi -  \varepsilon\) to \( \varepsilon\). Similar disturbances can occur when \(\delta_c\) jumps from \( \varepsilon\) to \(2\pi -  \varepsilon\).

We also run simulations when the bounded longitude phase \( \delta_c \) changes from \( \delta_c\) to \( \delta_c + \pi \) or \( \delta_c \) to \( \delta_c - \pi \). Owing to the latitude phase \( 2\alpha_c=\pi \) or \( 2\alpha_c=0 \)  when these jumps happen, there is no disturbance.

From the simulations above, we can see that the disturbances mainly occur when the longitude phase \( \delta_c \) jumps from \( 2\pi -\varepsilon \) to \( \varepsilon \) or from \( \varepsilon\) to \( 2\pi - \varepsilon\). The intensity of the disturbance depends on the latitude phase \( 2\alpha_c \). When the latitude phase approaches 0 or \( \pi \), corresponding to the north pole or south pole, the disturbance intensity approaches zero. When the latitude phase approaches \(  2\alpha_c =\pi/2 \), corresponding to the equator, the disturbance intensity approaches its maximum. To realize endless control and eliminate the disturbance, we design the endless control methods as follows.
\begin{figure}[!ht] 
	\centering\includegraphics[width=8.8cm]{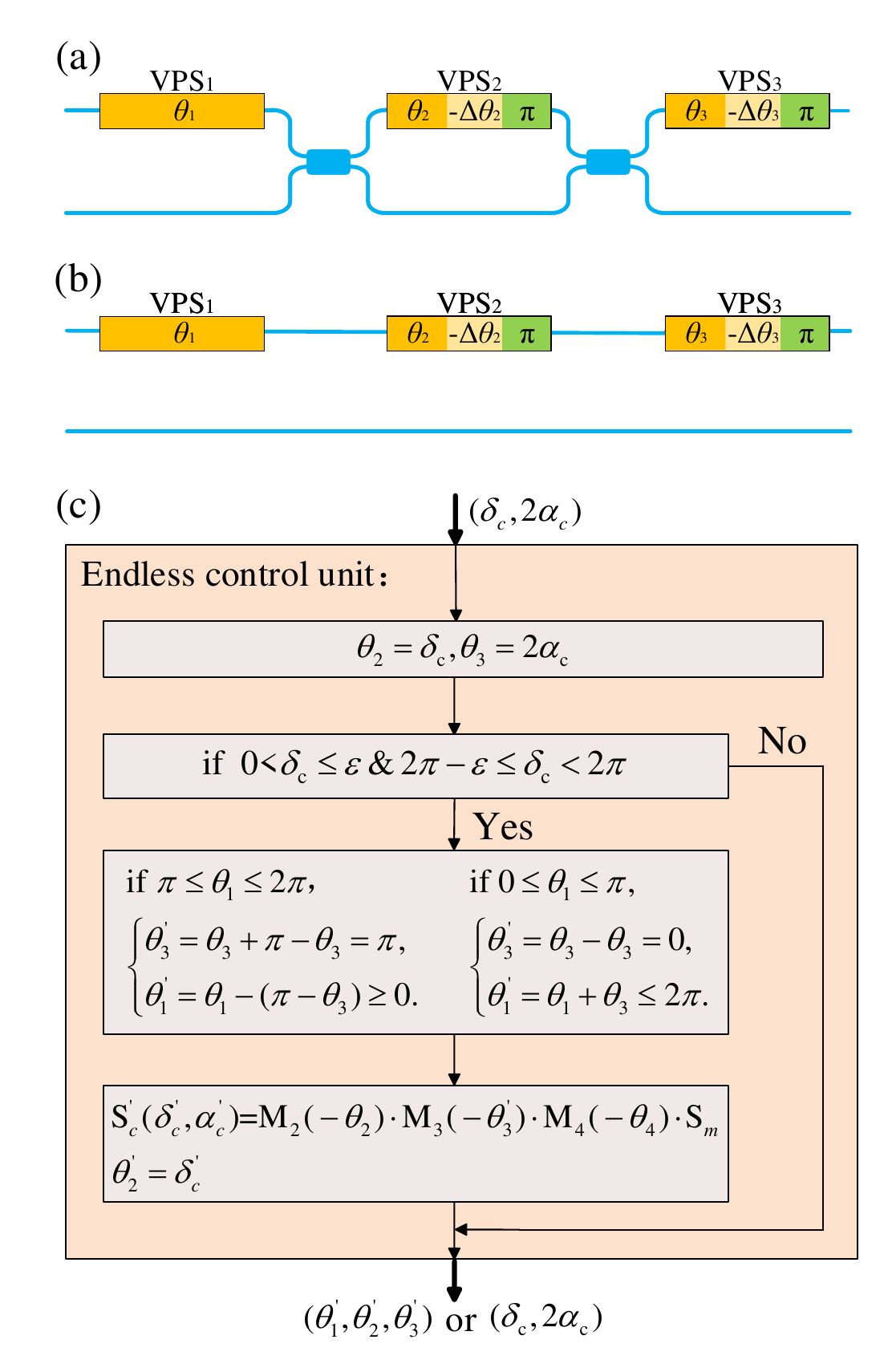}\\
	\caption{\label{9} Equivalent structure (a), (b) and flowchart (c) for the analytic polarization control.}
\end{figure}

In the normal control process, the phases \( \delta_c \) and \( 2\alpha_c \) are applied on the second control phase \( \theta_2 = \delta_c \) and the third control phase \( \theta_3 = 2\alpha_c \) respectively. When the derived control phase \( \delta_c \) approaches the boundary value 0 or \( 2\pi \) before a jump occurs, the first MZI structure as shown in Fig~\ref{9}(a) is equivalent to a directly connected structure as shown in Fig~\ref{9}(b) \cite{41}, and the control phases \( \theta_1 \) and \( \theta_3 \) are both in the equivalent upper waveguide. We use the operations in Eqs.~\ref{eq:27} and~\ref{eq:28}, as follows, to tune the first and third phases \( \theta_1 \) and \( \theta_3 \) without changing the output polarization state
\begin{align}
& \begin{cases}
    \begin{aligned}
    \theta'_3 &= \theta_3 + \pi - \theta_3 = \pi, \\
    \theta'_1 &= \theta_1 - (\pi - \theta_3) \geq 0,
    \end{aligned}
  \end{cases} 
  && \text{if } \pi \leq \theta_1 \leq 2\pi, \label{eq:27} 
  \tag{27}\\[10pt]
& \begin{cases}
    \begin{aligned}
    \theta'_3 &= \theta_3 - \theta_3 = 0, \\
    \theta'_1 &= \theta_1 + \theta_3 \leq 2\pi,
    \end{aligned}
  \end{cases} 
  && \text{if } 0 \leq \theta_1 \leq \pi. \label{eq:28}
  \tag{28}
\end{align}

In the above transformation, the control phases \( \theta_1 \) and \( \theta_3 \) vary in opposite directions, and the total phase in the upper waveguide remains constant. Thus, the output Stokes vectors remain unchanged. After the operation, the third control phase will be tuned to \( \theta_3' = 0 \) or \( \pi \).
The two operations are determined by evaluating the value of control phase \( \theta_1 \) to prevent \( \theta_1 \) from exceeding its boundary \([0, 2\pi]\). 

The control Stokes vector \( S_c(\delta_c, \alpha_c) \) will be changed to \(S_c'\), as calculated by the following equation
\begin{equation}
S_c'(\delta_c', \alpha_c') = M_2(-\theta_2) \cdot M_3(-\theta_3') \cdot M_4(-\theta_4) \cdot S_m.
\tag{29}
\label{eq:29}
\end{equation}

After the calculation, we should use \( \delta_c' \) and \( 2\alpha_c' \) to update the second and third control phases. It should be noted that \( S_m \) is at the north pole in polarization locking. After rotation around \( S_1 \) axis using \( M_4(-\theta_4) \), rotation around \( S_3 \) axis using \( M_3(-\theta_3') \), and rotation around \( S_1 \) axis using \( M_2(-\theta_2) \), the control Stokes vector \( S_c' \) is at the north or south pole. Thus, the updated latitude phase \( 2\alpha_c' = \theta_3' \) is 0 or \( \pi\). This means that the third control phase does not need to be updated in this step. Only the second control phase is updated from \( \theta_2 = \delta_c \) to \( \theta_2' = \delta_c' \). In this case, the longitude phase jumps from \( \delta_c \) to a new value \( \delta_c' \) instead of the jumping from \( 2\pi - \varepsilon\) to \(\varepsilon\) or from \( \varepsilon \) to \( 2\pi -\varepsilon\). This jump from \( \theta_2 = \delta_c\) to \( \theta_2'=\delta_c'\) will not introduce any disturbance because the third control phase or latitude phase, \( \theta_3' = 2\alpha_c' \) is 0 or \( \pi \), corresponding to the north or south pole. This phenomenon can also be interpreted using an equivalent structure. When the control phase \( \theta_3 \) is 0 or \( \pi \), the second MZI structure is equivalent to a direct or cross-connected connection structure. The jumps in the second control phase will not affect the output intensity of the upper waveguide.

In conclusion, the operation of Eqs.~\ref{eq:27}, \ref{eq:28} and ~\ref{eq:29} lies at the heart of endless control, which changes the control phases from \( \theta_1\), \(\theta_2 = \delta_c\), \(\theta_3 = 2\alpha_c \) to \(\theta_1'\), \(\theta_2' = \delta_c'\), \(\theta_3' = 2\alpha_c'\)
to avoid approaching the boundary and eliminate the intensity disturbance.

\section{Simulation of APC}
\subsection{Simulation results with and without endless control}
In this section, the overall process of analytic polarization control are presented. As shown in Fig.~\ref{10}, the flowchart can be divided into six steps, which will be introduced as follows:
\begin{figure}[!ht] 
	\centering\includegraphics[width=9cm]{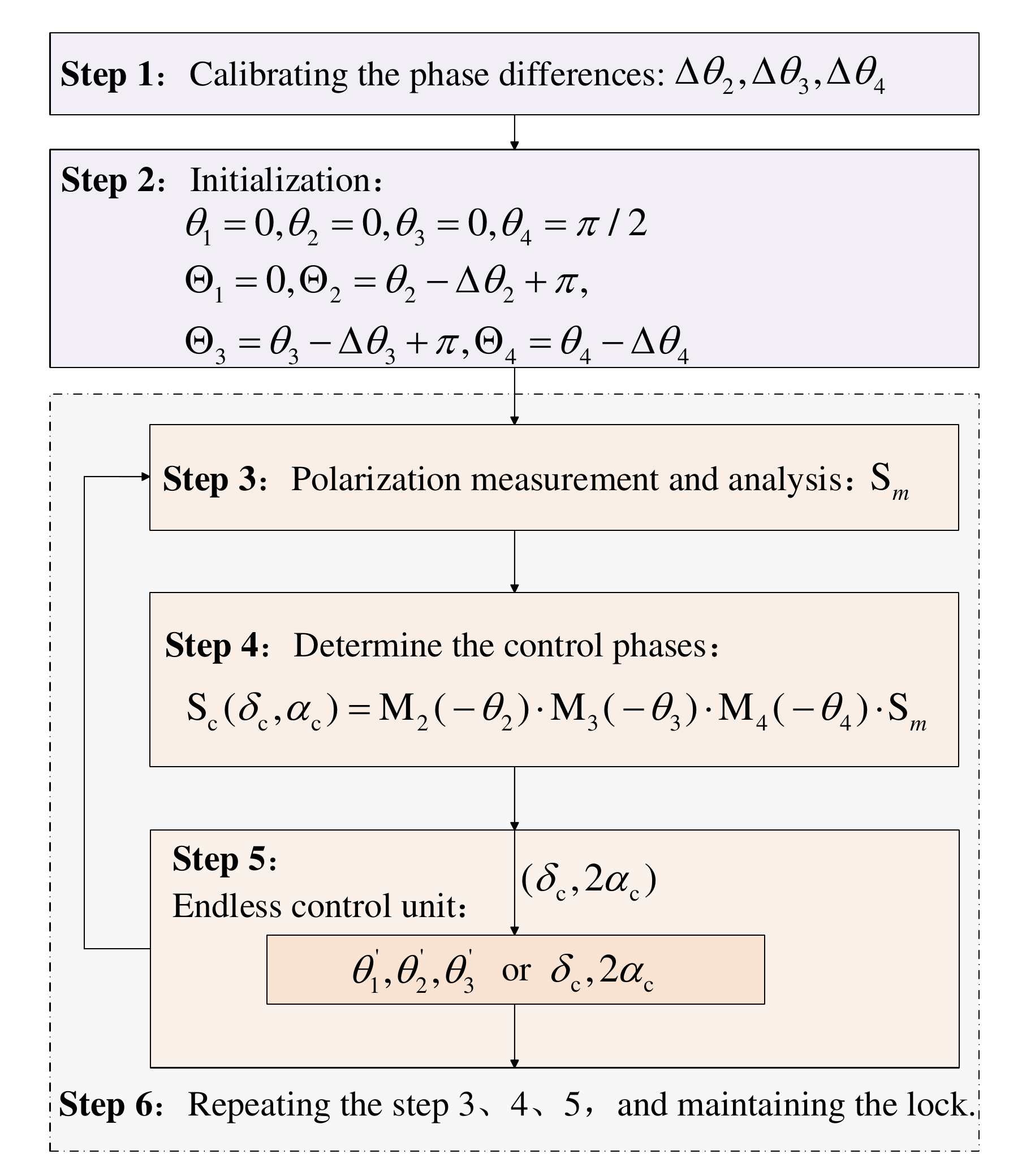}\\
	\caption{\label{10} Flowchart for APC.}
\end{figure}
\begin{figure}[!ht] 
	\centering\includegraphics[width=6cm]{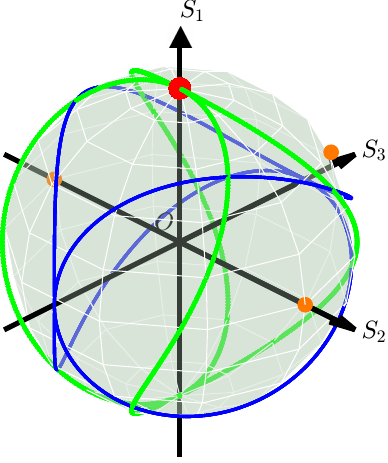}\\
	\caption{\label{11} The input, output, and control polarization states on the Poincaré sphere.}
\end{figure}

Step 1: Calibrate the phase differences \( \Delta\theta_2 \), \( \Delta\theta_3 \), and \( \Delta\theta_4 \) of each MZI structure.

Step 2: Initialize the control phase value in each control unit. Here, the values of these control phases are set to \( \theta_1 = 0 \), \( \theta_2 = 0 \), \( \theta_3 = 0 \), and \( \theta_4 = \pi/2 \). The total phase of each phase shift is set as \( \Theta_1 = \theta_1 \), \( \Theta_2 = \theta_2 - \Delta\theta_2 + \pi \), \( \Theta_3 = \theta_3 - \Delta\theta_3 + \pi \), and \(\Theta_4 = \theta_4 - \Delta \theta_4\). A nonzero initial value of the control phase \( \theta_4 \) is used to avoid the total phase \( \Theta_4 \) of the fourth phase shifter being a negative value.

Step 3: Measure the Stokes vector \( S_m \) in the measurement unit and analyze its phases.

Step 4: Derive the control Stokes vector \( S_c \) using Eq.~\ref{eq:18}, and analyze its phases \( \delta_c, \alpha_c \)

Step 5: Transform the control phases \( \delta_c, \alpha_c \) to new control phases \( \theta_1', \theta_2', \theta_3' \) or \( \theta_2, \theta_3\), and update these new control phases using the endless control unit.

Step 6: Repeat steps 3–5 and lock the polarization state to the desired polarization state.

Using the above mentioned steps, we can control any polarization state to the target polarization state using only one loop. Compared with the blind search-based method, which requires plenties of  loops, the analytic method saves a great deal of time and achieves the desired polarization state quickly.

Here, we simulate polarization control using the analytical method. We assume that phases \( \delta_{in} \), \( 2\alpha_{in} \) of input Stokes states \( S_{in} \) change at the same rate. In Fig.~\ref{11}, the green lines on the surface of the Poincaré sphere represent the varying input polarization states \( S_{in} \), and the blue lines represent the derived control polarization states \( S_{c} \). The red points at the north pole represent the output states \( S_{out} \), and the orange points represent the disturbed output polarization states without endless control.

\begin{figure}[!ht] 
	\centering\includegraphics[width=8.8cm]{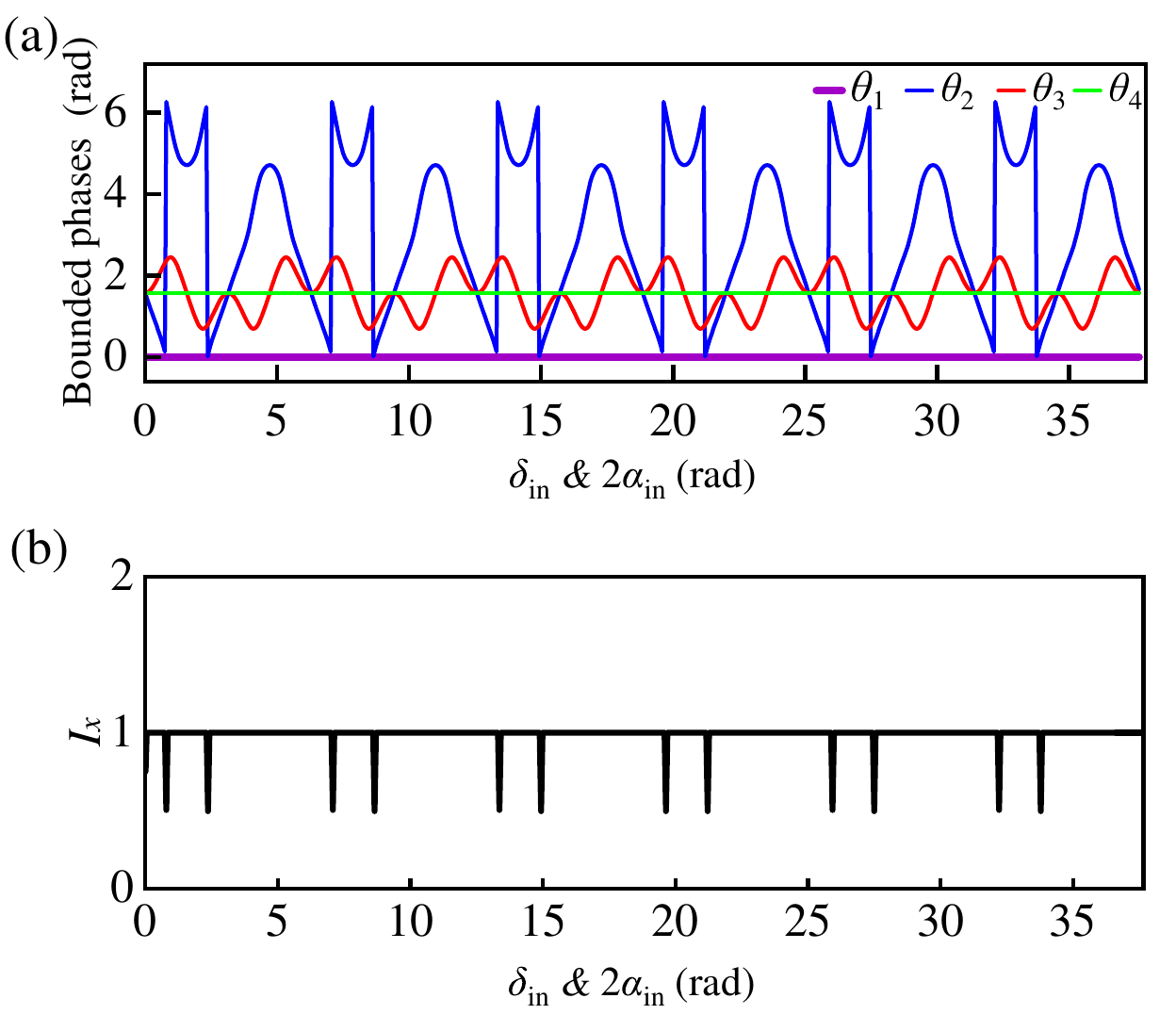}\\
	\caption{\label{12} Control phases (a) and output intensity $I_x$ (b) in APC without endless control.}
\end{figure}
Figure~\ref{12} shows the values of the control phases and the output intensity \( I_x \) in the upper waveguide in analytic polarization control without endless control. As the phases \( \delta_{in} \) and \( 2\alpha_{in}\) change, the control phases \( \theta_2 \) and \( \theta_3 \) vary accordingly. There are jumps in the control phases \( \theta_2 \), and there are no jumps in the control phases \( \theta_3 \). The control phases \( \theta_1, \theta_4 \) are constant throughout the whole process. Most of the time, the output intensity is constant with a normalized value of one. When the second control phase \( \theta_2 \) jumps from 0 to \( 2\pi \) or from \( 2\pi \) to 0, output intensity disturbance occurs.
\begin{figure}[!ht] 
	\centering\includegraphics[width=8.8cm]{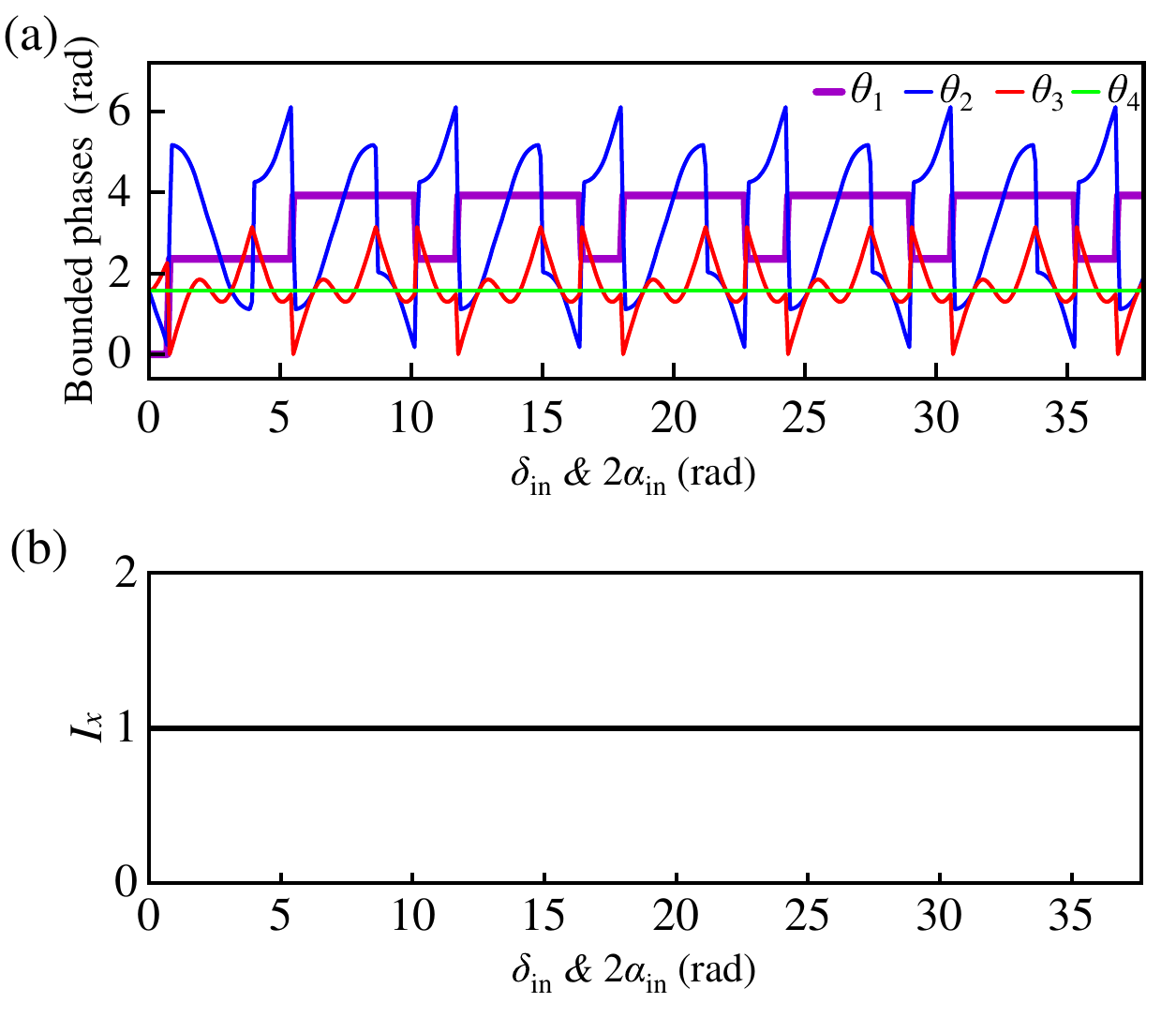}\\
	\caption{\label{13} Control phases (a) and output intensity $I_x$ (b) in APC with endless control.}
\end{figure}

Figure~\ref{13} shows the values of the control phases and the output intensity in the upper waveguide in APC with an endless control unit. As the phases \(\delta_{in} \) and \(2\alpha_{in}\) vary, when the control phase \( \theta_2 \) approaches to 0 or \( 2\pi \), the control phases \( \theta_1 \) and \( \theta_3 \) change simultaneously in opposite directions to \(\theta_1' \) and \( \theta_3' \), and the third control phase \( \theta_3' = 2\alpha_c' \) takes the value of 0 or \( \pi \). The first control phase remains in the range \([0, 2\pi]\). The second control phase \( \theta_2 \) jumps to a value \( \theta_2' = \delta_c' \) to stay away from boundary values. With the endless control unit, the output intensity of the upper waveguide \( I_x \) remains constant, indicating no disturbance in the output intensity. The fourth control phase \( \theta_4 \) remains constant value of \( \pi/2 \) throughout the entire process.

\subsection{Influence of the phase difference $\Delta\theta_4$  on the extinction ratio}

\begin{figure}[!ht] 
	\centering\includegraphics[width=8.8cm]{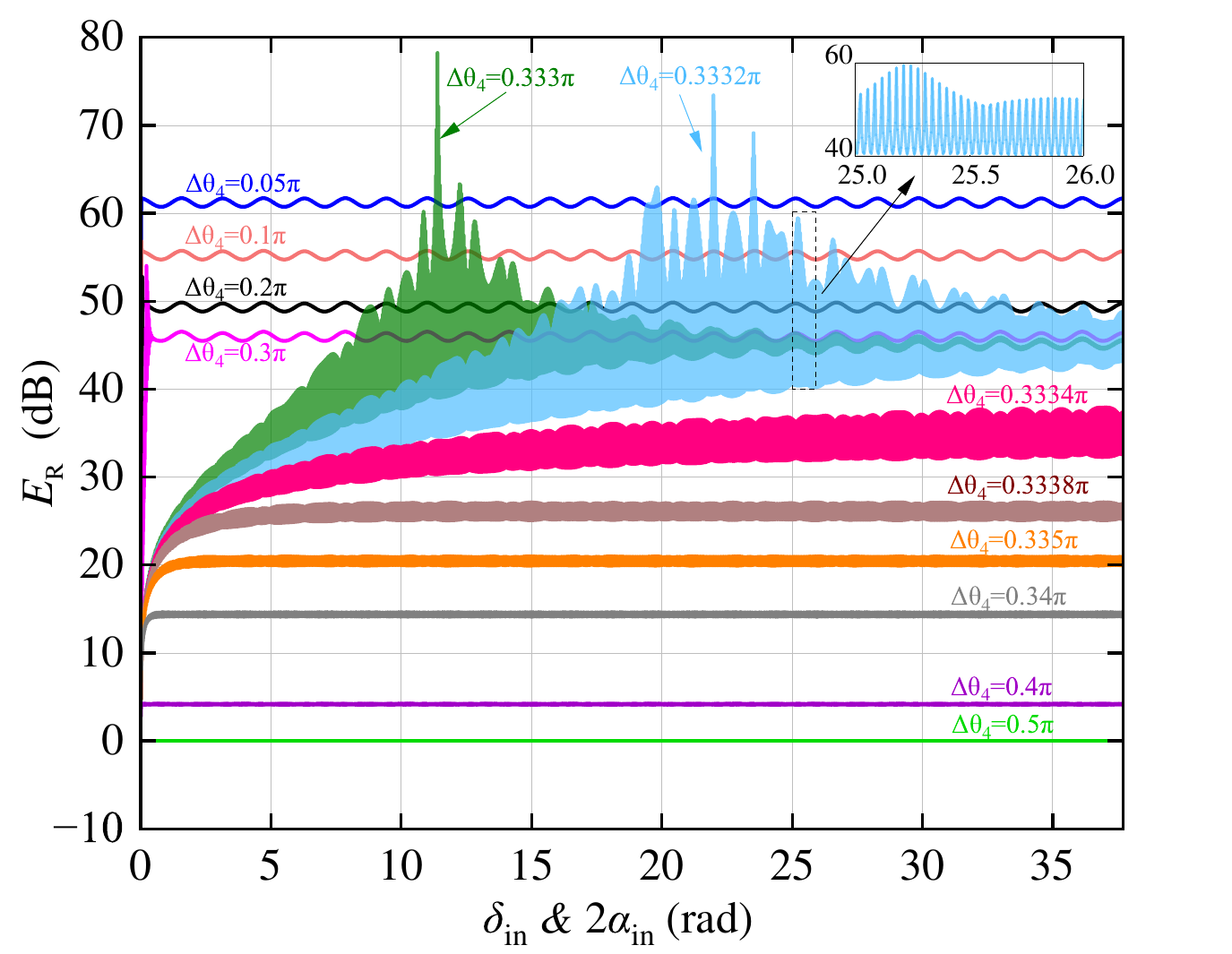}\\
	\caption{\label{16} Output extinction ratio versus the phases \( \delta_{in} \)  \& \( 2\alpha_{in} \) at different phase differences $\Delta \theta_4$.}
\end{figure}

From the above simulation results, we can see that the fourth control phase remains constant throughout the whole process. The fourth phase shifter is only used to compensate for the phase difference when rotations around \( S_1 \) and \( S_3 \) axes are utilized. To evaluate the effect of the fourth phase shifter, we perform a simulation of the endless APC without the fourth shifter. The extinction ratio for different phase difference \(\Delta \theta_4\) is presented in Fig.~\ref{16}. The extinction ratio \( E_R\) is defined as
\begin{equation}
E_R = 10 \cdot \log_{10} \left( \frac{I_x}{I_y} \right),
\tag{30}
\label{eq:30}
\end{equation}
where \( I_x \) and \( I_y \) denote the output intensities of the upper and lower waveguides, respectively. This extinction ratio can be used to evaluate the quality of the output Stokes vector. If the polarization is perfectly locked, the output beams are entirely in the upper waveguide, corresponding to the north pole. In this case, the extinction ratio is infinite. Owing to various factors, such as the extinction ratio of real devices, calibration errors, the split ratio errors of the 50/50 coupler, and uncompensated phase differences, the extinction ratio is restricted. Here, we focus on the effect induced by the phase difference.

From top to bottom, the phase differences of the twelve curves in Fig.~\ref{16} are \( 0.05\pi \), \( 0.1\pi \), \( 0.2\pi \), and \( 0.3\pi \), \( 0.333\pi \), \( 0.3332\pi \), \( 0.3334\pi \), \( 0.3338\pi \), \( 0.335\pi \), \( 0.34\pi \), \( 0.4\pi \), and \( 0.5\pi \), respectively. We can see that when the phase difference is less than \( 0.3\pi \), the output extinction ratio is greater than 40 dB. When the phase difference \( \Delta\theta_4 \) is between 0.3 and 0.335, the output extinction ratio is very sensitive to the phase difference. In this region, the extinction ratio decreases from  $\sim45$dB to $\sim15$ dB. The curves in this region widen because there are oscillations, as shown in the inset figure.

When the phase difference \( \Delta \theta_4 \) exceeds \( 0.335\pi \), the extinction ratio decreases to 20 dB or less. The extinction ratio \( E_R = 0 \) corresponds to the phase difference \( \Delta \theta_4 = \pi/2 \).

Here, only the results for positive phase differences are presented. When the phase differences are negative, similar results can be obtained if the absolute value of the phase difference is the same. Thus, to achieve a better extinction ratio in APC, it is recommended to use the fourth phase shifter to compensate for the phase difference.

\section{Conclusions}
Because the relative phase of the MZI structure can be controlled accurately on a photonics chip, we design and simulate an analytic polarization control method with four phase shifters. Three rotations around axes \( S_1 \), \( S_2 \), and \( S_3 \) and the corresponding structures for analytic polarization control are provided. The pairwise calibration method used to calibrate the phase difference of the MZI structure is refined and simulated. The endless control method is explored and simulated in detail. In the simulation of analytic polarization control, the values of various control phases, the output intensity, and their relationships are analyzed. Especially, the effect of the polarization control with and without an endless control unit are compared. The phase difference in measuring Stokes vectors can be compensated by the fourth phase shifter in the last or third MZI structure. To evaluate the importance of the fourth phase shifter, the output extinction ratio induced by the fourth phase difference is analyzed.

In our simulation, rotations around axes \( S_1 \) and \( S_3 \) are utilized. The operation of rotation around the \( S_2 \) axis can be realized by introducing the fourth phase shifter, and it opens the door to exploring more APC methods based on optical computing. The quality of the Stokes vector output depends mainly on the output extinction ratio, which in turn is influenced by multiple factors, such as the extinction ratios of actual devices, calibration inaccuracies, and the non-ideal splitting ratio of the 50/50 coupler. To achieve a better extinction ratio, the effect of these factors will be analyzed further in the future. More importantly, experimental validation of the proposed APC on a realistic four-phase-shifter photonics chip is anticipated, while FPGA-based acceleration is expected to enable high-speed dynamic polarization control with a high extinction ratio.

\par\vspace{14pt}
\noindent\textbf{Funding.} This work was supported in part by the National Natural Science Foundation of China under Grants No. 62671364, No. 62175138, No. 62205188, and
No. 11904219, and in part by the horizontal project of Beijing Xicheng Photonics Technology Co., Ltd under Grant No. 2026140105000302.

\vspace{14pt}
\noindent\textbf{Disclosures.} The authors declare no conflicts of interest.

\vspace{14pt}
\noindent\textbf{Data Availability.} Data underlying the results presented in this paper are not publicly available at this time but may be obtained from the authors upon reasonable request.

\providecommand{\noopsort}[1]{}
\providecommand{\singleletter}[1]{#1}


\end{document}